# Vertical Organic Thin-Film Transistor with Anodized Permeable Base for Very Low Leakage Current

*Felix Dollinger[+], Kyung-Geun Lim[+], Yang Li, Erjuan Guo, Peter Formánek, René Hübner, Axel Fischer, Hans Kleemann, Karl Leo\**

F. Dollinger, Y. Li, E. Guo, Dr. A. Fischer, Dr. H. Kleemann, Prof. K. Leo
Dresden Integrated Center for Applied Physics and Photonic Materials (IAPP), Technische Universität Dresden, 01062 Dresden, Germany
E-mail: felix.dollinger@tu-dresden.de, karl.leo@iapp.de

Dr. K.-G. Lim
Korea Research Institution for Standards and Science (KRISS),
267 Gajeong-ro, Yuseong-gu, Daejeon 34113, Republic of Korea
E-mail: kglim@kriss.re.kr

Dr. P. Formánek
Leibniz-Institut für Polymerforschung Dresden e. V. (IPF), 01069 Dresden, Germany

Dr. R. Hübner
Institute of Ion Beam Physics and Materials Research, Helmholtz-Zentrum Dresden-Rossendorf, 01328 Dresden, Germany

[+] First authors, equally contributing
\*[+] Corresponding authors

**Abstract**
The Organic Permeable Base Transistor (OPBT) is currently the fastest organic transistor with a transition frequency of 40 MHz. It relies on a thin aluminum base electrode to control the transistor current. This electrode is surrounded by a native oxide layer for passivation, currently created by oxidation in air. However, this process is not reliable and leads to large performance variations between samples, slow production and relatively high leakage currents. Here, we demonstrate for the first time that electrochemical anodization can be conveniently employed for the fabrication of high performance OPBTs with vastly reduced leakage currents and more controlled process parameters. Very large transmission factors of 99.9996% are achieved, while excellent on/off ratios of $5 \times 10^5$ and high on-currents greater than 300 mA/cm² show that the $C_{60}$ semiconductor layer can withstand the electrochemical anodization. These results make anodization an intriguing option for innovative organic transistor design.

**Introduction**

The performance and processability of organic thin-film transistors (OTFTs) have been considerably improved in recent years. Nowadays, OTFTs are readily employed on flexible substrates, and operate in the MHz regime.[1-6] In particular, the transition frequency $f_t$ is an important figure of merit for the transistor performance because it weights the transconductance over the device capacitance and hence is a measure for the effectiveness of conductance per area. The present record for $f_t$ of organic transistors is 40 MHz, enabled by a vertical transistor concept – the Organic Permeable Base Transistor (OPBT).[7-9] Its vertical conductive channel facilitates the transistor to carry on-current densities as high as 1 kA/cm².[8] Furthermore, utilizing the full overlap area of the electrodes simultaneously results in a minimal parasitic capacitance and ultimately into record-high $f_t$.[7] Hence, the OPBT is a promising device concept that bears the potential for further device- and process-innovation in order to realize even better organic transistors in the future.

The OPBT is built up in a vertical stack with $C_{60}$ as semiconductor material, sandwiched between two electrodes (denoted as emitter and collector) and separated in the middle by a thin metal electrode, called the base electrode (cf. Figure 1). The 15 nm thin base electrode allows electrons to pass through nanometer-sized pinholes while the device is in its on-state, and blocks electrons in its off-state.[10] The thin base electrode layer, generally made of vacuum-deposited aluminum, grows in grains on the organic material.[11] Pinholes at grain-boundaries of the metal film facilitate charge carrier transport across the electrode. The transmission of electrons is modulated effectively by an electrical potential applied between emitter and base $V_{BE}$, resulting in a high on/off-ratio of the transistor. An aluminum-oxide layer around the base electrode metal is vital to suppress leakage current $I_B$ into the base electrode and to enable a high transmission factor $\alpha$, being the ratio of transmitted current $I_T \approx I_{Collector} \equiv I_C$ to the total emitter current $I_E$.[12] According to Equation (1), $\alpha$ increases linearly with decreasing $I_B$

$$\alpha = \frac{I_C}{I_E} = \frac{I_C}{I_B + I_C} = 1 - \frac{I_B}{I_B + I_C} \approx 1 - \frac{I_B}{I_C} \text{ for } I_B \ll I_C \ . \tag{1}$$

In state-of-the-art OPBTs, the oxide layer is formed by a brief exposure of the device to ambient air during the deposition process.[9,13-16] This procedure has been demonstrated to result in transmission factors of $\alpha$ = 99.00%,[17] meaning that $I_B$ is only two orders of magnitude lower than $I_C$ which is unacceptable for circuits with low static power consumption.[18] However, the same procedure has been employed to demonstrate hero devices with $\alpha$ = 99.99%.[9] Hence, despite its appealing simplicity, this fabrication procedure has strong limitations concerning leakage currents and device reproducibility which ultimately prevent OPBTs from being used in complex electronic circuits.

In this work, we show a substantial improvement of the passivating oxide film of the base electrode by wet electrochemical anodization. We control the electrochemical growth of the oxide and tune the dielectric properties of the ultra-thin base electrode. As a result, a tremendous reduction in base-leakage current is achieved. While other groups have successfully used anodization to improve the performance of lateral $C_{60}$ based thin-film transistors,[19,20] they have not exposed the semiconductor itself to the electrochemical process. We show for the first time that it is possible to use such a process on $C_{60}$ while preserving the device function and a high current density. Overall, the formation of tight and robust oxide films by electrochemical oxidation renders the possibility to fabricate high-performing OPBTs with ultra-low leakage currents with a defined and reliable process.

**Results and Discussion**
**Figure 1** shows an idealized device structure of a vertical OPBT with an anodized permeable base electrode. The formation of the anodized base oxide layer depicted in the schematic image is precisely controlled by the anodization parameters. In order to form this thin AlOx layer, the half-fabricated OPBT (the bottom (collector) electrode, the bottom $C_{60}$ layer, and the base electrode) is potentiostatically anodized in citric acid electrolyte as documented in **Figure S1** and described in the Experimental Section. The anodization occurs at the interface between elementary aluminum and the electrolyte, thereby passivating the full Al surface with a nonporous barrier-type aluminum oxide.[21] Directly after anodization, the optical appearance of the anodized device has changed due to the formation of the transparent AlOx layer, and the coloring of the structure provides a good indicator for the film thickness (cf. Figure S1 d).

The impact of anodizing conditions on device characteristics of OPBTs with electrochemically oxidized base electrodes is investigated (**Figure 2**). Quite impressively, the device structure, including the organic semiconductor material, is not severely damaged during the wet anodization process. All devices show a reliable transistor behavior with on-currents greater than 300 mA/cm² at a driving voltage of 1 V, which are comparable to OPBTs fabricated by dry oxidation.[17] In the transfer curves of the transistors, the on-currents and the leakage currents (base currents) decrease with increasing anodizing potential (Figure 2a). For anodizing potentials of 2 V and 4 V, the base leakage current of OPBTs remains on the lowest level within the measurement limit for large parts of the transfer curve. The device with an anodization voltage of 2 V shows a transmission factor of 99.9996% at equal collector-emitter and base-emitter voltages of 1 V, corresponding to a current gain of $2.5 \cdot 10^5$. The base leakage current ($10^{-10}$ A) is decreased by about 4 orders of magnitude compared to a naturally oxidized base in previous reports,[9,17] with the leakage current being suppressed below the measurement limit for an anodization voltage of 4 V. With the base current being that low and decoupled from the collector and emitter currents, the OPBT can now be regarded as a field effect-transistor, as already proposed by simulations.[10] Although, the transmission is increased due to the anodization, the on-current of the transistor is reduced for higher anodizing voltage. This effect probably originates from a lower electric base-emitter field across the oxide in combination with a reduced number of pinholes (discussed later in the manuscript).

With increasing anodizing voltage and hence oxide layer thickness, the capacitance of the base oxide is reduced. Figure 2b shows a decrease in capacitance (solid line) between base and emitter $C_{BE}$ with higher anodizing potential. Due to the scattering of data (caused by variation in the electrode overlap, cf. Figure S1d), the thickness of the AlOx layer cannot reliable be deduced from the capacitance value in the accumulation regime (positive base-emitter voltage). However, the phase of the impedance signal (dotted lines) at a constant frequency of 1 kHz remains close to -90° (ideal capacitor) for anodizing potentials of 2 V and 4 V, which proves the excellent blocking of the base leakage current by the tight AlOx film formed by wet anodization. The sample anodized at 2 V shows an increase in leakage current starting at a base-emitter voltage of 1.2 V (Fig. 2a). Simulations reveal that this happens exactly when charge carrier accumulation starts at the bottom of the base layer.[10] Hence, it can be concluded that the anodization at 2 V creates a very dense oxide layer on top of the base layer that successfully blocks all leakage current.

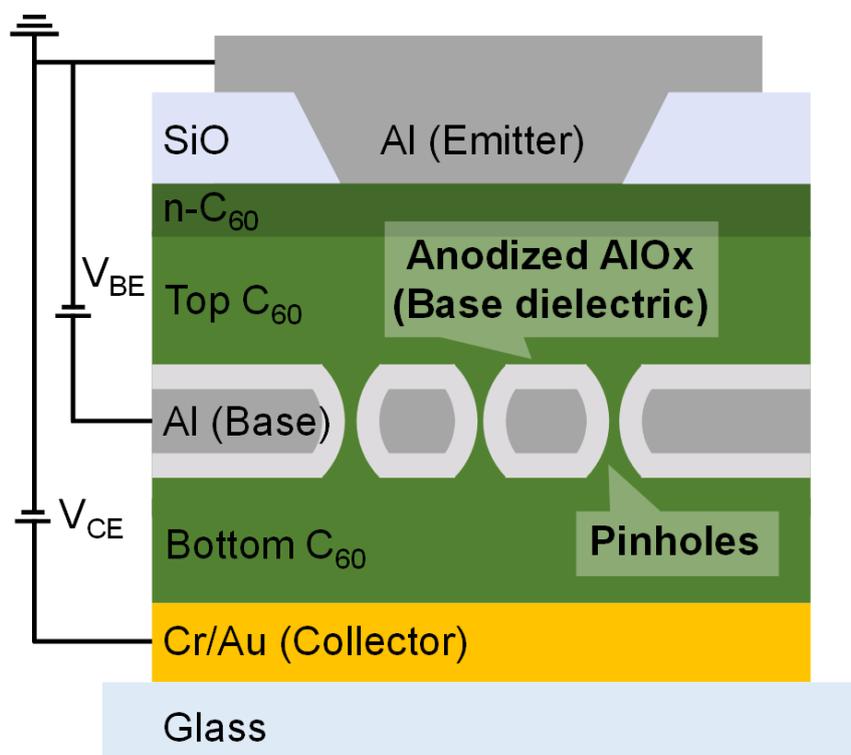

*Figure 1 Device schematics of a vertical OPBT with an electrochemically oxidized permeable base electrode. The half-fabricated OPBT (glass, chromium/gold electrode, bottom $C_{60}$ layer, and base aluminum) is subjected to electrochemical oxidation (anodization). The surrounding surface of the base is potentiostatically anodized to form a thin barrier type aluminum oxide. After anodization and drying, the top $C_{60}$ film, the $C_{60}$ layer n-doped with $W_2(hpp)_4$, the SiO structuring layer, and the emitter aluminum layer are sequentially evaporated.*

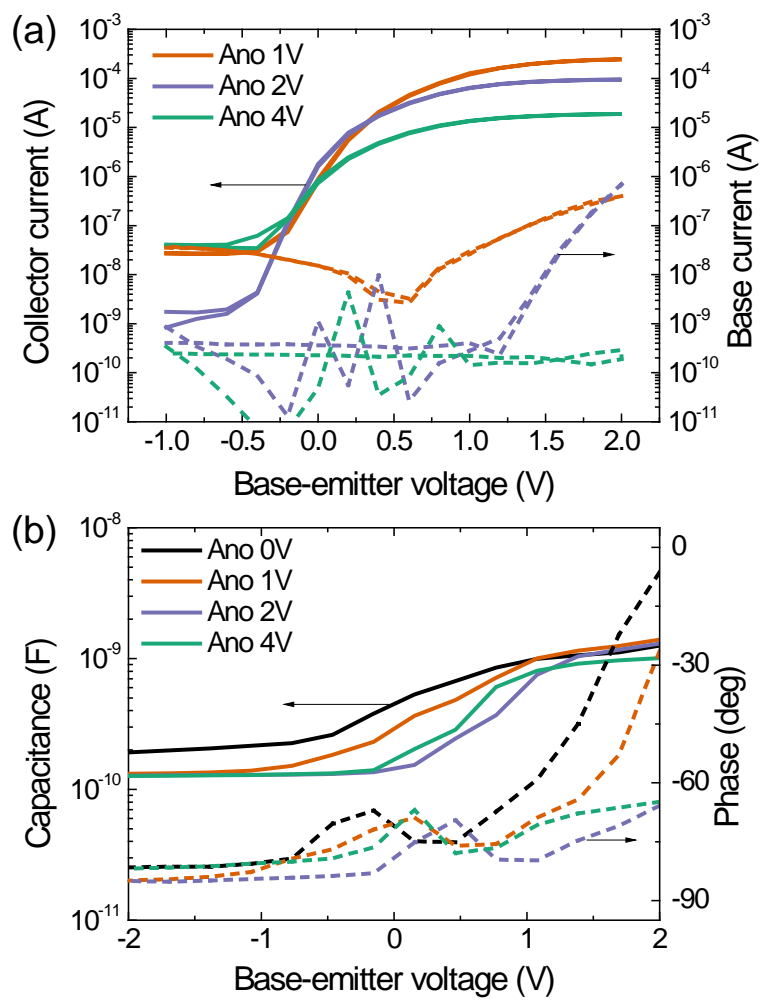

**Figure 2** *OPBTs with an electrochemically oxidized permeable base electrode. The Al base is potentiostatically anodized with anodizing potentials of 1 V, 2 V, and 4 V.* **(a)** *Transfer curves with an emitter-collector voltage $V_{CE}$ of 1 V. Both, on-current and base leakage current decrease with anodizing potential, but the decrease in leakage current is much larger.* **(b)** *Capacitance and phase curves between base and emitter with a constant frequency of 1 kHz. The phase in anodized samples stays close to -90°, indicating excellent insulating properties of the base oxide layer.*

Overall, the electrochemically oxidized OPBTs show on-state current densities comparable to state-of-the-art OPBTs fabricated by dry oxidization, however, their leakage current is significantly reduced. Most remarkably, this has been achieved with a wet chemical oxidation process atop of an air-sensitive organic semiconductor material which to the best of our knowledge is presented for the first time in this manuscript.

In order to get experimental access to the thickness of the AlO$_x$ layer, we analyze element distributions obtained by spectrum imaging based on EDXS analysis in STEM mode from cross-sectional TEM lamellae prepared from different representative devices. **Figure 3** shows cross-section of a OPBTs with dry oxidation and with anodization at 2 V and at 4 V. The measured oxide thickness increases with the anodizing potential from 5 nm (no anodization, only oxidation in ambient), to 6 nm (2 V anodization) and 10 nm at 4 V anodization, which proves that the thickness of the oxide film can be controlled precisely by the anodizing voltage.

Using the previously published OPBT layer structure with a 15 nm thin base electrode,[22] an anodizing voltage of 4 V would correspond to a residual thickness of the Al electrode of only 5 nm, which might lead to a severe and undesired drop of its conductivity. For this reason, we evaluate the anodization technique for devices with thicker base layers (**Figure S2**). While devices with ambient oxidation barely shows any noticeable transmission with a 50 nm thick Al base electrode, anodized samples clearly work as transistors with this thick Al layer, which means that pinholes through the base layer are present and we suspect the pinholes to be created by strain induced by the anodization process. This difference in the behavior between OPBTs fabricated by anodization and dry oxidation suggests that the microstructure of the base electrode is different for both methods. Hence, in the following, we discuss the microstructure in more detail.

The surface morphology of AlO$_x$ layers from different oxidation processes is investigated by SEM, as can be seen in **Figure S3**, comparing ambient oxidation and anodization at 2 V and 4 V. The contrast between grains is relatively high with dry oxidation and low with anodization voltage. Because $O_2$ molecules in ambient air thermally diffuse to Al grain boundaries, the AlOx grows irregularly with dry oxidation.[23] However, OH- ions in the anodizing electrolyte drift to the Al surface driven by a static potential which is uniform on the surface. Thus, AlOx grows evenly on the surface with anodization. Therefore, the surface of pinholes, which are mainly located at grain boundaries, is also evenly oxidized during anodization. Another difference in the surface properties which can be seen in Figure S3, is that the Al grains tend to become larger with anodization while the density and size of the pinholes is reduced. This observation is consistent with the finding that the collector current is reduced with increasing anodization potential (cf. Figure 2). We speculate that differences in the magnitude and distribution of strain in the base layer during the oxidation step (anodization vs. ambient oxidation) gives rise to the lower pinhole density for anodized samples.

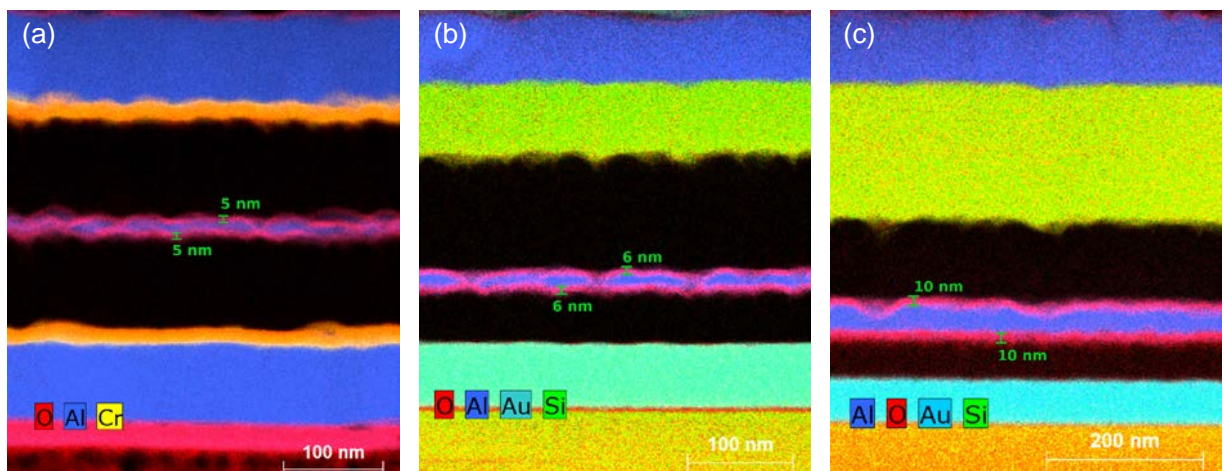

*Figure 3 Elemental distribution in energy dispersive X-ray spectroscopy (EDX) analysis in TEM using electron transparent cross-sections of OPBTs **(a)** with a 15 nm base layer oxidized in ambient air, **(b)** with a 15 nm base layer anodized at 2 V, **(c)** with a 50 nm base layer anodized at 4 V. A clear increase in oxide layer thickness can be observed. The oxide thickness is the same above and below the base layer. The SiO layers in (b) and (c) are used for structuring as depicted in Figure 1 a. They are not present in the active area. Coloring of emitter and collector differs, because different metals have been used.*

A final aspect about the anodization that should be discussed is the conformity of the anodized film. The $AlO_x$ layer is formed not only at the top surface but also at the surrounding surface and inside the nanometer-sized pinholes of the Al base electrode, passivating the entire metal film against current leakage from or into the semiconductor. However, even if the cross-sectional TEM suggests within its resolution limit an equal thickness of the oxide at the bottom and top surface of the base electrode (cf. Figure 3), it is a priori not clear to be an inherent property of the anodization process. In order to investigate this aspect more thoroughly, we analyze the capacitance between base and collector ($C_{BC}$) as a measure of the $AlO_x$ thickness underneath the base. The values for $C_{BC}$ increase with the thickness of the Al base layer (cf. Figure S2 a, for negative voltage), which means in turn that the thickness of the anodized $AlO_x$ layer underneath the base electrode decreases with a thicker base electrode. This is because the infiltration of the electrolyte through the nanometer-sized pinholes of the base electrode is stochastically blocked with increasing thickness of the base electrode. $C_{BC}$ of $AlO_x$ underneath a 15 nm Al base is doubled compared to a 50 nm thick Al base with an identical anodizing potential of 1 V, while the capacitance between emitter and base ($C_{EB}$) is independent of the thickness of the base electrode (Figure S2 b). The bottom side of the base layer facing the collector is less exposed to the anodization, because it is passivated by the bottom $C_{60}$ layer. Moreover, although the anodization enhances the oxide layer thickness also on the bottom side of the base layer, the lowered phase of the impedance data for the collector-base capacitor clearly shows that the oxidation is less perfect, because the electrolyte may not reach every part of the metal-$C_{60}$ interface.

**Conclusion**

In conclusion, vertical OPBTs with a large on/off ratio (> $10^5$), large on-current-density (> 300 mA/cm²), and record transmission factors of 99.9996% are demonstrated by employing a wet electrochemically oxidized permeable base electrode. The Al base electrode located in the middle of the vertical OPBT device is exposed to an electrolyte and anodized to form an $AlO_x$ surface around the base layer and inside the nanometer-sized pinholes. In contrast to naturally oxidized $AlO_x$ layers, the dielectric properties of anodized $AlO_x$ layers underneath and at the top surface of the base electrode are controlled by anodizing conditions and the thickness of the Al base electrode. Our results show that anodization is a viable method to create an excellent and well defined dielectric oxide for OPBTs and that organic semiconductors can withstand this process.

This work sets the grounds for the reliable fabrication of high-performance OPBTs. The improved degree of fabrication reliability together with the low static power loss / low base leakage current might enable the integration of OPBTs in complex electronic circuits in the future. Moreover, the method proposed here allows for controlling the device capacitance and employing thicker base metal films resulting in a lower electrode resistance. Both aspect are of utmost importance for the development of even faster organic transistors.

**Experimental Section**

For anodization, half-devices are prepared by vacuum evaporation on a thoroughly cleaned glass substrate. A thin chromium film (3 nm) improves adhesion of the gold electrode (50 nm), upon which a $C_{60}$ layer (50 nm) and the base aluminum (varying thickness) are evaporated under ultra-high-vacuum conditions through a set of shadow masks. Then the sample is removed from the vacuum chamber and stored in inert $N_2$ atmosphere, except for a duration of about 1 hour when the anodization is performed. In order to improve the interfacial adhesion between the semiconductor layer and the base electrode we employ a heating step prior to the anodization. The half-fabricated OPBT samples without upper semiconductor layer and emitter electrode are preheated at 150°C for 1 hour before anodizing the base electrode in order improve the contact area and interfacial adhesion of $C_{60}$ and base. As a result, $C_{BC}$ of OPBTs with preheating shows larger charge accumulation and phase shift compared to without preheating and the on/off ratio and base leakage are significantly improved (**Figure S4**). Prior to the anodization, a nail protection polymeric coating is employed outside of the active device area (Figure S1a) in order to avoid damage to the electrodes from potential peaks at the water level. Anodization is done in a solution of 1 mM/L citric acid (2-hydroxypropane-1,2,3-tricarboxylic acid) in deionized water. Voltages (1 V to 4 V) are applied with respect to an aluminum counter-electrode using a Keithley 2400 Source Measure Unit (SMU), as shown in Figure S1 b. The anodization current is shown in Figure S1 c. The current is applied until a stable plateau is reached after 15 to 60 seconds. The effect of the anodization can be seen by naked eye in a color change of the thin Al metal film (Figure S1 d). After anodization, the samples are left to dry in medium vacuum conditions for 30 minutes and then exposed to heat-treatment on a hotplate in $N_2$ atmosphere. Thereupon, the samples are returned to the vacuum deposition chamber ($< 10^{-6}$ mbar), where the remaining layers are evaporated. On top of the base, a second $C_{60}$ film is evaporated with a larger thickness of 100 nm. A layer of highly efficient n-dopant $W_2(hpp)_4$ (20 nm, 1 wt.% in $C_{60}$) is used for improved electron injection.[24,25] A layer of SiO (100 nm) is structured by shadow masks to limit the contact area of the top electrode (Al, 100 nm) to the doped semiconductor to 250 µm x 250 µm, thereby defining the active area. Pinholes in the aluminum base layer form natively.[22] After fabrication in the vacuum chamber, the samples are encapsulated under $N_2$ atmosphere (< 1 ppm $O_2$ and $H_2O$) with cavity glasses and epoxy glue. The thickness of the base oxide layer depends on the anodization condition. A thicker oxide layer reduces the amount of elemental aluminum remaining in the layer. To characterize the microstructure and chemical composition of the OPBTs, cross-sectional transmission electron microscopy (TEM) analysis is performed. For that purpose electron transparent specimens are prepared by in-situ lift-out using a NEON40 focused ion beam (FIB) device (Carl Zeiss Microscopy GmbH, Germany). Transmission electron microscopy measurements are carried out with a Libra200 (Carl Zeiss Microscopy GmbH, Germany) operated at an acceleration voltage of 200 kV. High-angle annular dark-field scanning transmission electron microscopy and spectrum imaging based on energy dispersive X-ray spectroscopy (HAADF-STEM and EDXS) are conducted with a Talos F200X (Thermo Fischer Scientific / FEI, USA) operated at 200 kV and equipped with a Super-X EDX detector. Electrical characterization of the devices is done with a Keithley SCS 4200 Parameter Analyzer in ambient air, or with a Keithley 2400 and a Keithley 2611a SMU in a glovebox. For the impedance spectroscopy an HP 4284A LCR-Meter is used.

**Acknowledgements**
F. Dollinger and K.-G. Lim contributed equally to this work.
Funding by the Deutsche Forschungsgemeinschaft (DFG) within the FlexART project LE 747/52-1 in the DFG Priority Program SPP 1796 is gratefully acknowledged, as well as within the Cluster of Excellence Center for Advancing Electronics Dresden (cfaed) and the EFOD project RE 3198/6-1. This work has supported by the National Research Foundation of Korea (NRF) grant funded by the Korea government (MSIT) (2018R1C1B6004221). The use of HZDR Ion Beam Center TEM facilities and the funding of TEM Talos by the German Federal Ministry of Education of Research (BMBF), Grant No. 03SF0451 in the framework of HEMCP are acknowledged.
Thanks and acknowledgement go to Michael Göbel (IPF) for FIB lamella preparation and Tobias Günther and Andreas Wendel (both IAPP) for sample fabrication.

**References**
[1] K. Fukuda, Y. Takeda, Y. Yoshimura, R. Shiwaku, L. T. Tran, T. Sekine, M. Mizukami, D. Kumaki, S. Tokito, *Nature Communications,* **2014**, 5, 4147, DOI: 10.1038/ncomms5147

[2] A. Perinot, P. Kshirsagar, M. A. Malvindi, P. P. Pompa, R. Fiammengo, M.Caironi, *Scientific Reports,* **2016**, 6, 38941, DOI: 10.1038/srep38941
[3] V. Fiore, P. Battiato, S. Abdinia, S. Jacobs, I. Chartier, R. Coppard, G. Klink, E. Cantatore, E. Ragonese, G. Palmisano, *IEEE Transactions on Circuits and Systems I: Regular Papers,* **2015**, 62, 1668–1677, DOI: 10.1109/tcsi.2015.2415175
[4] A. Yamamura, H. Matsui, M. Uno, N. Isahaya, Y. Tanaka, M. Kudo, M. Ito, C. Mitsui, T. Okamoto, J. Takeya, *Advanced Electronic Materials,* **2017**, 3, 1600456, DOI: 10.1002/aelm.201600456
[5] U. Kraft, T. Zaki, F. Letzkus, J. N. Burghartz, E. Weber, B. Murmann, H. Klauk, *Advanced Electronic Materials*, **2018**, 0, 1800453, DOI: 10.1002/aelm.201800453
[6] J.W. Borchert, U. Zschieschang, F. Letzkus, M. Giorgio, M. Caironi, J. N. Burghartz, S. Ludwigs, H. Klauk, *IEEE International Electron Devices Meeting (IEDM)*, **2018**, 38.4.1-38.4.4, DOI: 10.1109/IEDM.2018.8614641
[7] B. Kheradmand-Boroujeni, M. P. Klinger, A. Fischer, H. Kleemann, K. Leo, F. Ellinger, *Scientific Reports*, **2018**, 8, 7643, DOI: 10.1038/s41598-018-26008-0
[8] M. P. Klinger, A. Fischer, F. Kaschura, J. Widmer, B. Kheradmand-Boroujeni, F. Ellinger, K. Leo, *Scientific Reports,* **2017**, 7, 44713, DOI: 10.1038/srep44713
[9] M. P. Klinger, A. Fischer, H. Kleemann, K. Leo, *Scientific Reports*, **2018**, 8, DOI: 10.1038/s41598-018-27689-3
[10] F. Kaschura, A. Fischer, M. P. Klinger, D. H. Doan, T. Koprucki, A. Glitzky, D. Kasemann, J. Widmer, K. Leo, *Journal of Applied Physics,* **2016**, 120, 094501, DOI: 10.1063/1.4962009
[11] K.-G. Lim, M.-R. Choi, J.-H. Kim, G. H. Jung, Y. Park, J.-L. Lee, T.-W. Lee, *ChemSusChem*, **2014**, 7, 1125-1132, DOI: 10.1002/cssc.201301152
[12] A. Fischer, P. Siebeneicher, H. Kleemann, K. Leo, B. Lüssem, *Journal of Applied Physics*, **2012**, 111, 044507, DOI: 10.1063/1.3686744
[13] Al-Shadeedi, *PhD Thesis*, Kent State University, May, **2017**, URL: https://etd.ohiolink.edu/!etd.send_file?accession=kent1492441683969202
[14] K. Agrawal, O. Rana, N. Singh, R. Srivastava, S. S. Rajput, *Applied Physics Letters*, **2016**, 109, 163301, DOI: 10.1063/1.4964838
[15] G. Giusi, E. Sarnelli, M. Barra, A. Cassinese, G. Scandurra, K. Nakayama, C. Ciofi, *IEEE Transactions on Electron Devices*, **2017**, 64, 4260-4265 DOI: 0.1109/TED.2017.2738699
[16] K.-I. Nakayama, S.-Y. Fujimoto, M. Yokoyama, *Organic Electronics*, **2009**, 10, 543-546, DOI: 10.1016/j.orgel.2009.02.003
[17] M.P. Klinger, A. Fischer, F. Kaschura, R. Scholz, B. Lüssem, B. Kheradmand-Boroujeni, F. Ellinger, D. Kasemann, K. Leo, *Advanced Materials*, **2015**, 27, 7734-7739, DOI: 10.1002/adma.201502788
[18] A. Al-Shadeedi, S. Liu, V. Kaphle, C.-M. Keum, B. Lüssem, *Advanced Electronic Materials*, 0, 1800728, DOI: 10.1002/aelm.201800728

[19] Z. Jianlin, C. Rengang, *Journal of Semiconductors*, **2011**, 32, 024006, DOI: 10.1088/1674-4926/32/2/024006

[20] G. Schwabegger, M. Ullah, M. Irimia-Vladu, M. Baumgartner, Y. Kanbur, R. Ahmed, P. Stadler, S. Bauer, N. Sariciftci, H. Sitter, *Synthetic Metals*, **2011**, 161, 2058-2062, DOI: 10.1016/j.synthmet.2011.06.042

[21] M. Kaltenbrunner, P. Stadler, R. Schwödiauer, A. W. Hassel, N. S. Sariciftci, S. Bauer, *Advanced Materials*, **2011**, 23, 4892-4896, DOI: 10.1002/adma.201103189

[22] A. Fischer, R. Scholz, K. Leo, B. Lüssem, *Applied Physics Letters*, **2012**, 101, 213303, DOI: 10.1063/1.4767391

[23] P. Dumas, J. Dubarry-Barbe, D. Rivière, Yves Levy, J. Corset, *Journal de Physique Colloques*, **1983**, 44, pp.C10-205-C10-208, DOI: 10.1051/jphyscol:19831042

[24] A. A. Günther, M. Sawatzki, P. Formánek, D. Kasemann, K. Leo, *Advanced Functional Materials*, **2016**, 26, 768-775, DOI: 10.1002/adfm.201504377

[25] B. Lüssem, C.-M. Keum, D. Kasemann, B. Naab, Z. Bao, K. Leo, *Chemical Reviews*, **2016**, 116, 13714, DOI: 10.1021/acs.chemrev.6b00329

Supporting Information

**Vertical Organic Thin Film Transistor with Anodized Permeable Base for Very Low Leakage Current**

*Felix Dollinger, Kyung-Geun Lim, Yang Li, Erjuan Guo, Peter Formánek, René Hübner, Axel Fischer, Hans Kleemann, Karl Leo*

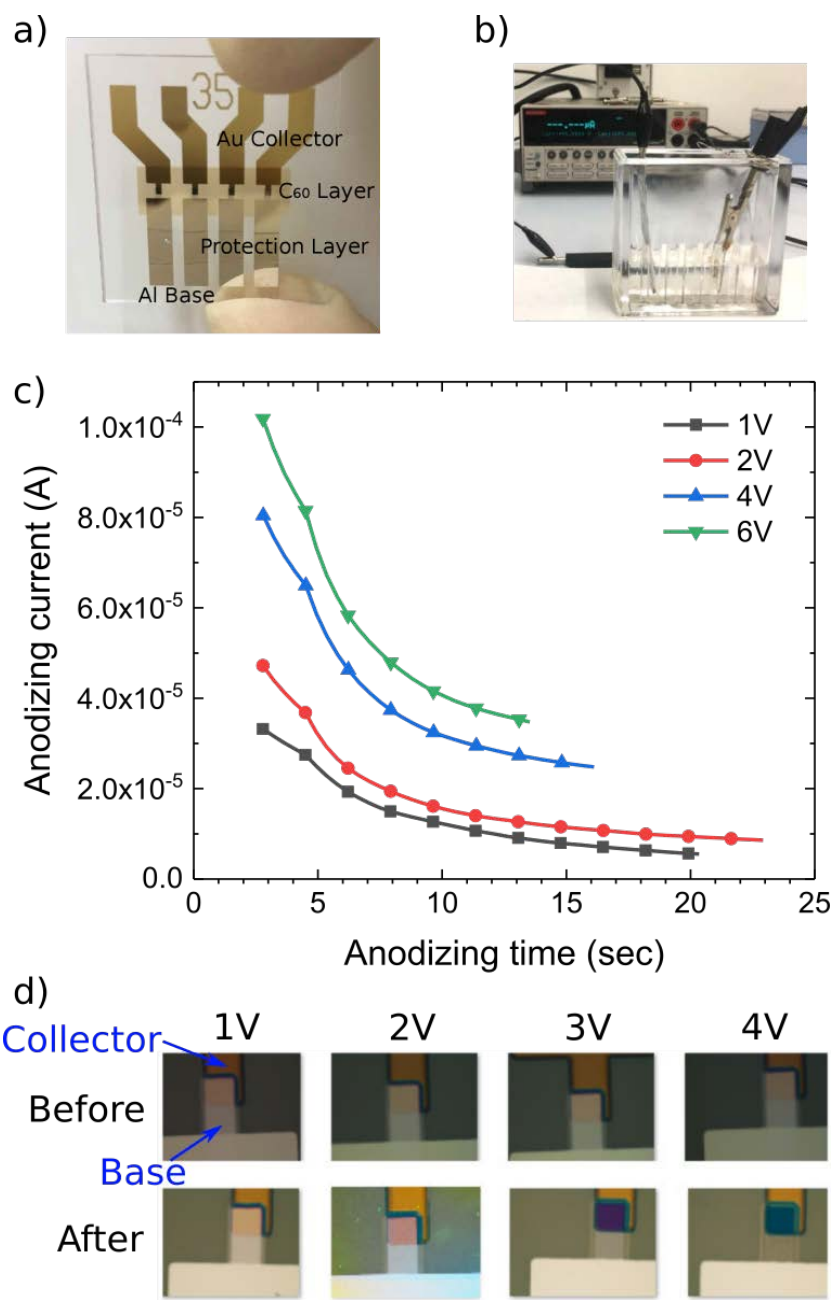

**Figure S1 (a)** Photograph of a substrate with four samples prepared for anodization, showing the collector electrode (Au), the bottom semiconductor layer ($C_{60}$) and the base electrode (Al). A polymer film has been applied to the collector contact to avoid damage at the water level in the anodization bath. **(b)** Sample in the anodization bath. The base electrode is contacted with a clamp and the anodization voltage is applied with respect to a metallic counter-electrode in the electrolyte. **(c)** Drop of the anodization current with time at various voltages. As the Al electrode gets passivated by a native oxide layer, the current level decreases and reaches a plateau. The absolute level of the currents depends on the particular geometry of the anodization bath. **(d)** Micrographs of samples before and after anodization. The contact on the top-side is the collector electrode, while the bottom one is the thin base electrode. A change in apparent color shows an increase in oxide thickness with anodization potential.

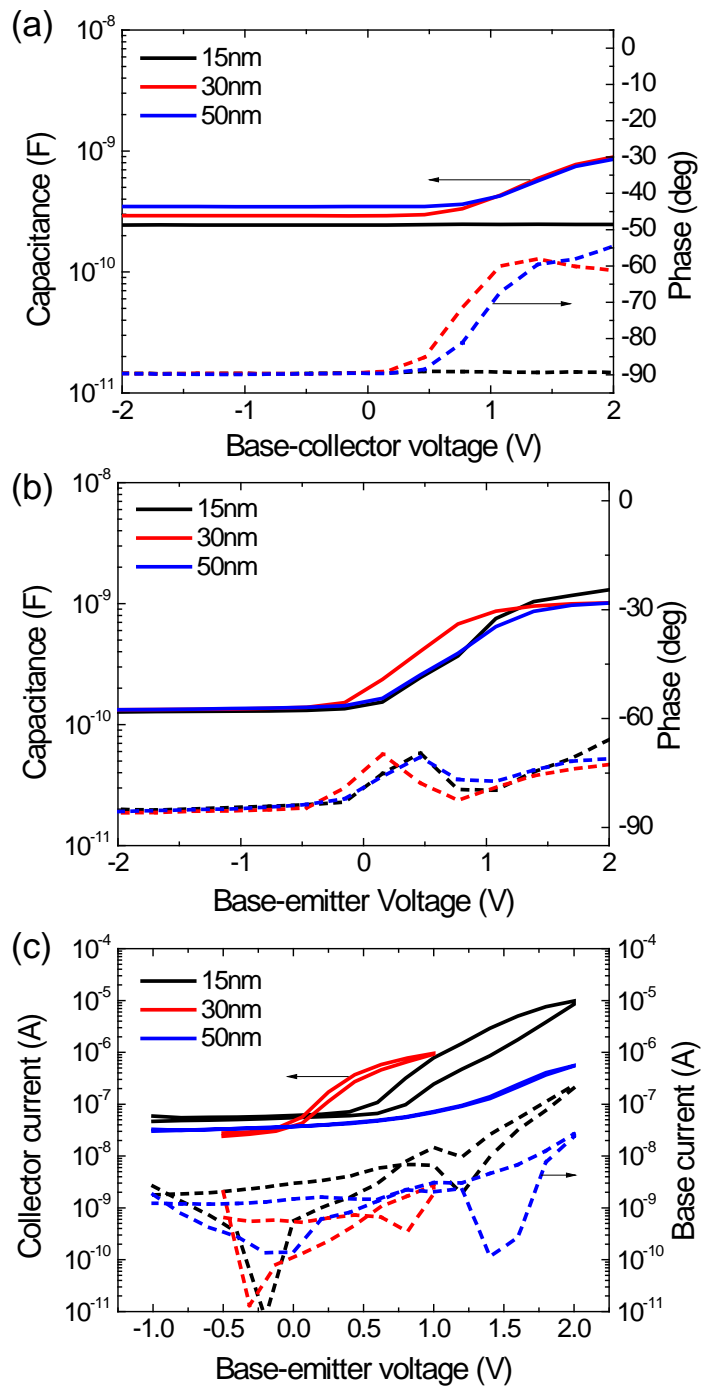

**Figure S2** The dielectric properties and the OPBT characteristics of anodized AlOx layers with various thickness underneath and atop of the base electrode. The capacitance and the phase curves with constant frequency of 1 kHz between (a) Base-Collector (underneath surface of base) and (b) Emitter-Base (atop surface of base) of OPBTs with 15, 30, and 50 nm-thick base electrode. (c) Transfer curve with $V_{CE}$ of 1 V for the OPBTs. Anodizing potential is 1 V for (a) and 2 V for (b) and (c), respectively.

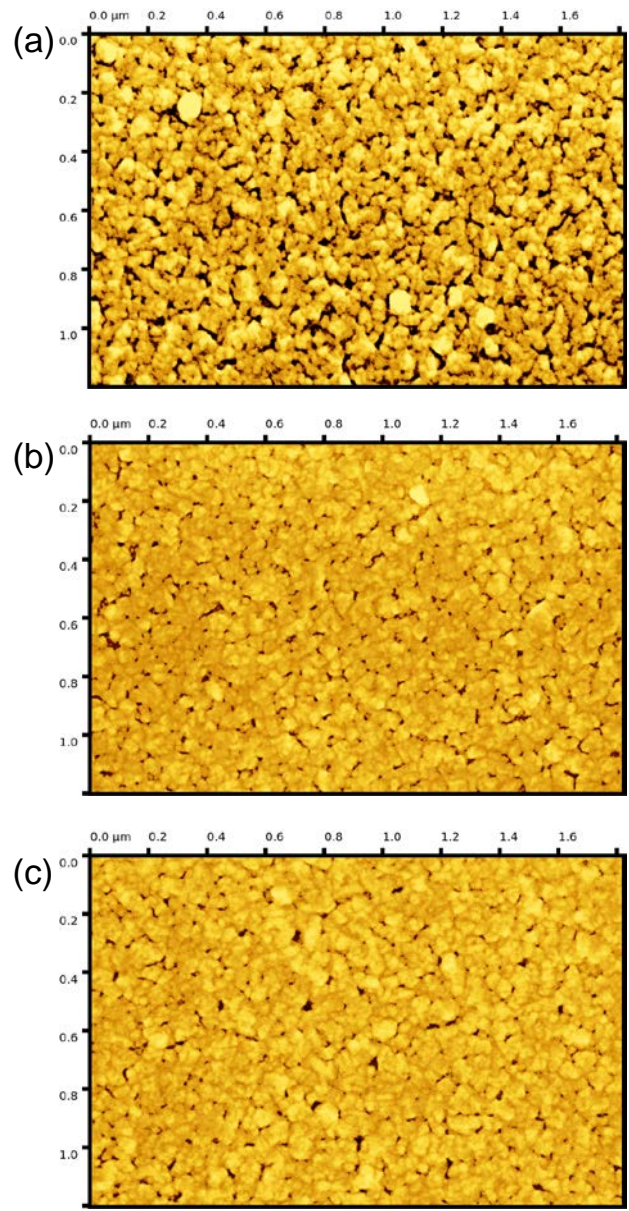

**Figure S3** SEM images of the base layers (a) with ambient oxidation, (b) after anodization at 2 V, and (c) after anodization at 4 V. The aluminum layers were evaporated onto a gold electrode and a 50 nm $C_{60}$ layer to ensure comparable growth conditions. Grains are highlighted with a yellow image mask using a threshold filter. It can be seen that the grain size of the Al/AlOx film increases with anodization, while the pinhole density is reduced. Automated grain counts result in the following data: a) 488 grains per µm², average grain area $2.0 \times 10^{15}$ m², b) 34 grains per µm², average grain area $26.9 \times 10^{15}$ m², c) 41 grains per µm², average grain area $24.1 \times 10^{15}$ m².

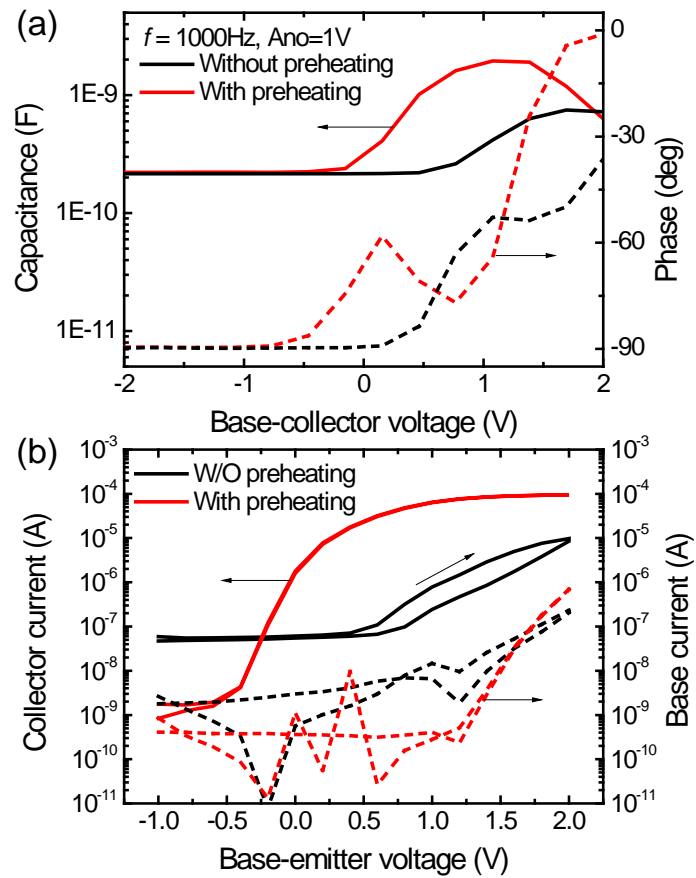

**Figure S4** Electronic characteristics of OPBTs with preheating of the permeable base. (a) The capacitance and the phase curves between base-collector with constant frequency 1 kHz and (b) transfer curve with $V_{CE}$ = 1 V of OPBTs with a heat treatment of 150 °C for 1 hour in the middle of the fabrication process before anodization.

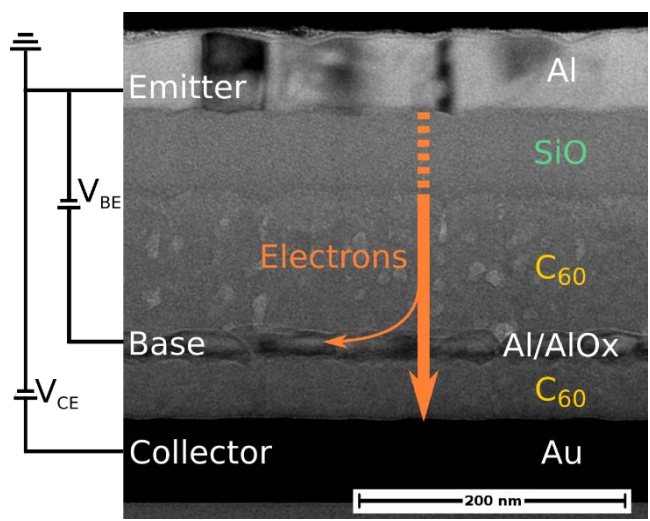

**Figure S5** Applied voltages and electron flow illustrated on a cross-sectional transmission electron microscope (TEM) image of an anodized OPBT. A large current is injected from the emitter electrode and transported through the vertical device by the bulk $C_{60}$. In contrast to a field-effect transistor, where a thin conductive channel is formed, the entire electrode area of the OPBT is used for injection. Most of the current is transferred to the collector electrode (described by the transmission factor α), while some fraction is lost into the base electrode as leakage current. The SiO insulator layer visible in this cross-sectional image is not present in the active area of the transistor. It is applied to insulate the emitter electrode outside of the active area.